\documentstyle[12pt,preprint,lscape]{aastex}
\newcommand{\etal}{et~al.}
\received{2003 January 22}
\begin{document}

\title{
The Early Light Curve of the Optical Afterglow of GRB 021211 
}

\author{Weidong Li, Alexei V. Filippenko,
Ryan Chornock, and Saurabh Jha \\
Email: (wli, alex, rchornock, sjha)@astro.berkeley.edu}

\affil{Department of Astronomy, University of California, Berkeley,
CA 94720-3411} 

\slugcomment{Submitted to ApJ Letters}

\begin{abstract}

The Katzman Automatic Imaging Telescope (KAIT) at Lick Observatory observed GRB
021211 starting 105~s after the burst, among the earliest observations of an
optical afterglow to date. Our well-sampled light curve, consisting of 18
points in the first 10 min and a total of 28 useful observations in the first
2.5~hr after the GRB, directly shows for the first time a break at $t \approx
10$ min.  The light curve can be represented as a sum of two components with
power-law decay indices of $-1.82\pm0.02$ and $-0.82\pm0.11$.  We hypothesize
that the data before the break were dominated by emission from the reverse
shock, as was previously suggested for GRB 990123.  Our data suggest that
either GRB 021211 underwent a dramatic color change at early times, or there
are small-scale variations superimposed on the power-law decay of the
reverse-shock emission.  The faintness of GRB 021211, coupled with the fast
decline typical of optical afterglows, suggests that some of the ``dark bursts"
were not detected because optical observations commenced too late after the
GRB.

\end{abstract}

\keywords{gamma rays: bursts}

\section{INTRODUCTION}

The majority of gamma-ray burst (GRB) optical afterglow (OA) observations
have been obtained several hours after the bursts, owing mainly to the delay in
getting X-ray images and/or the difficulties in localizing the GRBs in real
time. Out of the $\sim 40$ detected GRB OAs, only GRB 990123 (Akerlof et
al. 1999), GRB 021004 (Fox 2002), and GRB 021211 (Fox \& Price 2002) were
successfully detected within 10 min after the initial burst of gamma rays.

In this {\it Letter}, we report on early-time optical observations of GRB
021211 with the Katzman Automatic Imaging Telescope (KAIT).  KAIT is a 0.76-m
robotic telescope at Lick Observatory that is dedicated to searching for and
observing supernovae and monitoring other celestial events. It is equipped with
an Apogee AP7 back-illuminated CCD camera having a resolution of $0\farcs8$
pixel$^{-1}$ and a total field of view of $\sim$ $6\farcm8\times6\farcm8$. More
information on KAIT can be found in Li et al. (2000) and Filippenko et
al. (2001), and the KAIT GRB alert system is described in detail elsewhere (Li
et al. 2003).

\section{OBSERVATIONS}

On 2002 December 11, 11:18:34.03 UT (defined as $t=0$ in this paper), a bright
GRB triggered the HETE-2 satellite (Crew et al. 2002). The gamma-ray light
curve shows a single peak lasting $\sim$ 2.5~s. The flight localization of the
GRB was reported in a GRB Coordinates Network (GCN; Barthelmy \etal~1994)
notice at $t = 22$~s.

The KAIT GRB alert program received the GCN notice with the position of GRB
021211 at $t = 32$~s. At the time, KAIT was conducting followup observations of
one of its supernova discoveries, SN 2002he, and was 478~s into a 600-s
$B$-band exposure (a rare, long exposure; most are only 20~s). KAIT performed
some checks to ensure that the GRB was observable, stopped the autoguider,
rotated the filter wheel so that no filter was in the way of the light path,
and began to slew across much of the sky to the GRB position at $t =
49$~s. Currently, KAIT cannot stop an exposure in progress, so the shutter
remained open during the long move to the opposite side of the sky, fortunately
imaging only the dark, slowly rotating dome.  The dome slit was centered on the
GRB position at $t = 108$~s, with 48~s left in the ongoing exposure. The
effective start time for the GRB exposure is $t = 105$~s.\footnote{We have
conducted tests to repeat the same pointing procedure, and concluded that the
dome cleared the telescope 3~s earlier than centering on the GRB position.}
This exposure, with the SN 2002he field and the GRB 021211 field superimposed,
is shown in Figure 1a.

KAIT then began its pre-programmed sequence of GRB images. It started with a
10-s exposure at $t = 171$~s (Fig. 1b), continued with a 40-s exposure, five $3
\times 20$~s grid images (for a total of 15 individual exposures of the GRB),
two 60-s exposures, one 120-s guided exposure, one 300-s guided exposure, and
four 120-s guided exposures that together make a mosaic covering
$12\farcm7\times12\farcm7$ around the GRB position. One of the $3 \times 20$~s
grid images is shown in Figure 1c. Such grid images, in which the telescope is
moved by $36''$ in right ascension between consecutive exposures, save two
readout times (10~s each).

KAIT received three more GCN notices on GRB 021211 in the next 3 hr, and
repeated the same sequence each time. The notices arrived at 11:53:30 UT,
12:25:24 UT, and 13:29:33 UT, which correspond to 2096~s, 4010~s, and 7859~s
after the burst, respectively.  KAIT was conducting the regular supernova
search (20-s exposures) when these notices arrived, and was able to start the
respective GRB exposure 35~s, 23~s, and 45~s after receiving the GCN notice.  A
total of 61 images (including the one superimposed on the field of SN 2002he)
were obtained, with 101 individual exposures of GRB 021211.  Unfortunately,
there were passing clouds during the night, and the short exposures in the last
three sequences do not go sufficiently deep to detect the GRB.

Since our observations were obtained without filters, it is important to
transform the measured magnitudes to a standard photometric system, so that
comparisons with observations obtained elsewhere can be made.  Details of
transforming the unfiltered magnitudes to the standard Cousins $R$ are
described by Li et al. (2003); see also Riess et al. (1999). These papers
demonstrate that when the unfiltered observations are treated as though they
were obtained with a broadband filter, and color terms are applied to correct
for the difference in the throughput curves, the observed unfiltered magnitude
can be transferred to a standard passband with a precision of $\sim$ 5\%. The
color terms are determined by observing Landolt standard stars in $BVRI$ and
without filters during photometric nights; a value of 0.27$\pm$0.05 mag is
found for transforming KAIT unfiltered magnitudes to Cousins $R$. Knowledge of
the ($V - R$) color of GRB 021211 is also needed for the transformation, and we
assume ($V - R$) = 0.2 mag, similar to that of GRB 990123 (Fruchter et
al. 1999). A different choice of ($V - R$) for GRB 021211 will make our
measurements systematically brighter or fainter. For example, if ($V - R$) =
0.5 mag were used, our measurements would need to be made 0.08 mag brighter.

We used the PSF-fitting technique to reduce our data. For the grid images, the
PSF for an individual exposure is constructed by using stars in that exposure
only. Absolute photometric calibration of the GRB 021211 field is provided by
Henden (2002). Our final photometry in the $R$ band is listed in Table 1. In
total, 28 useful data points were measured from the observations, 18 of which
were obtained within 10 min after the burst.

\section{RESULTS}

Figure 2 displays the early-time light curve of GRB 021211 from the KAIT
observations. We assume no host galaxy reddening to GRB 021211, and adopt a
Galactic extinction of $A_R = 0.065$ mag (Schlegel, Finkbeiner, \& Davis 1998)
for the figure.  The data show a steep initial decline, well fit by a power-law
decay, with indexes measured as $-1.56\pm0.02$ for the first 4 points, and
$-1.58\pm0.02$ for the first 12 points. An apparent break is seen at $t \approx
0.1$--0.2 hr, when the GRB switches to a slower power-law decay, with indexes
measured as $-1.01\pm0.05$ from the last 7 points and $-0.94\pm0.06$ from the
last 6 points. The power-law indexes reported here are consistent with, but
supersede, those measured by Chornock et al. (2002) from the preliminary
reduction of the KAIT data (Li et al. 2002).

To better describe the temporal evolution of the OA of GRB 021211, we fit the
data with the sum of two power-law decays: 

\begin{equation}
 F_\nu = 0.5 \, F_{\nu,b}\, [(t/t_b)^{-\alpha_1} + 
(t/t_b)^{-\alpha_2}],  
\end{equation}

\noindent 
where $t_b$ is the time of the break, and $F_{\nu,b}$ is the flux at $t_b$.  We
obtained the following values for the parameters using a $\chi^2$-minimizing
procedure: $\alpha_1$ = 1.82$\pm$0.02, $\alpha_2$ = 0.82$\pm$0.11, and $t_b$ =
0.18$\pm$0.03 hr (= 10.2 min).  These parameters are consistent with those
found by Fox et al. (2003) with independent later-epoch data.  The solid line
in the upper panel of Figure 2 shows this fit, while the lower panel shows the
residuals of the fit.  The reduced $\chi^2$ of the fit is 3.34 for 25 degrees
of freedom, indicating that eq. (1) is not a good model for the data at the
8$\sigma$ confidence level,\footnote{We have additionally tried to fit the data
with smoothly broken power-law models of varying sharpness, but these did not
change the results significantly.} and suggest that the short timescale
variations in the residuals seen in the lower panel of Figure 2 (and especially
before $t_b$) are real features.

If we exclude the first data point, the model fit to eq. (1) becomes
acceptable (reduced $\chi^2$ = 1.03). However, we have no {\it a priori} reason
to suspect the first data point is anomalous. For the first point to be
consistent with the simple double-power-law model, (1) the flux-weighted mean
exposure time would have to be 135.0~s, which is 4.5~s after the midpoint of the
exposure, (2) the photometry error bar would have to be increased to 0.08
mag, or (3) the GRB would have to have been 0.11 mag brighter than our
measurement. The first option requires the GRB to be brightening during
the exposure, and is ruled out as the earlier RAPTOR image (Wozniak et al.
2002) shows the GRB already declining. The second option is also not
viable, because the OA is bright and well-detected. The third option is
possible if the color of the GRB underwent a dramatic change at early
times --- e.g., the $V-R$ color was 0.6 mag during the first
exposure, and became bluer to 0.2 mag (adopted in the photometric
reductions) starting from the second exposure.

\section{DISCUSSION}

The OA of GRB 021211 was first identified by Fox \& Price (2002) in images
taken with the Palomar 1.2-m Oschin Schmidt telescope, beginning about 20 min
after the burst. It was caught at early times by two robotic telescopes besides
KAIT.  The RAPTOR team obtained one image of GRB 021211 at $t$ = 65~s (Wozniak
et al. 2002), the second-earliest detection ever of an afterglow (only GRB
990123 was detected earlier at $t$ = 23~s; Akerlof \etal~1999), and Super-LOTIS
observations began at $t$ = 143~s (Park, Williams, \& Barthelmy 2002).
Attempts to detect the afterglow of GRB 021211 in other energy bands all
failed: these include the prompt GeV/TeV emission by Milagro (McEnery 2002),
the sub-mm search by Hoge et al. (2002) on Dec. 12, and the radio searches by Berger
\& Frail (2002) on Dec. 12 and by Rol \& Strom (2002) on Dec. 12 and 17.
Late-time spectroscopy at the location of GRB 021211 shows underlying emission
consistent with a host galaxy at probable redshift 1.006 (Vreeswijk et
al. 2003).

GRB 021211 is only the second GRB with prompt optical data, so it is of
interest to compare its light curve with that of GRB 990123 (Akerlof et
al. 1999). In Figure 3, the data for GRB 990123 were collected from various
sources as listed in Fruchter et al. (1999); only the last 5 ROTSE points are
included. We have also adopted the time zero point for GRB 990123 from 
Fruchter et al. (1999). The data for GRB 021211 reported in the GCN Circulars
are also plotted. We fit the same power-law model [eq. (1)] to the GRB
990123 data within 2~d after the burst and obtained the following parameters:
$\alpha_1$ = 1.80$\pm$0.11, $\alpha_2$ = 1.06$\pm$0.05, and
$t_b$ = 0.15$\pm$0.04 hr (= 9.0 min). The reduced $\chi^2$ of the fit
is 1.10 for 20 degrees of freedom, which indicates eq. (1) is an adequate
model to fit the data of GRB 990123. The fit parameters of GRB 990123 are very
similar to those of GRB 021211,  except for perhaps $\alpha_2$, as the
late-time decline of GRB 990123 is a bit steeper (as shown also by the
residuals in the lower panel of Figure 3). 
Although the temporal evolution of both GRBs
is consistent with a break at $t \approx $ 10 min, the well-sampled KAIT data
of GRB 021211 are the first to directly demonstrate the break and show a smooth
transition between the two power-law decays. 

Despite the overall similarity between the temporal evolution of the OA of the
two bursts, GRB 990123 is brighter than GRB 021211 by about 3 mag at similar
epochs, even though the former is probably at a larger redshift ($z$ = 1.600;
Hjorth et al. 1999). Other important differences between the two bursts are
that (1) GRB 990123 is a much longer burst (lasting $\sim$ 63.3~s; Galama et
al.  1999) than GRB 021211 (lasting only $\sim$ 2.5~s); (2) GRB 990123 is
exceptionally bright in gamma rays, and has a fluence that is in the top
0.3\% of BATSE bursts, while the gamma-ray fluence of GRB 021211 is in the 
middle of HETE-2 bursts reported by Barraud et al. (2003); and (3) the afterglow
of GRB 990123 was successfully observed in optical, infrared, sub-mm, mm, and
radio passbands, while the afterglow of GRB 021211 is detected only in the
optical and infrared. In particular, a radio flare was observed for GRB 990123
(Kulkarni et al. 1999; Galama et al. 1999) and interpreted by Kulkarni et
al. (1999) as evidence of reverse-shock emission.  The apparent faintness of
the OA of GRB 021211 compared with the OA of GRB 990123 suggests that the
afterglow of GRB 021211 in other energy bands may be fainter as well,
contributing to the reasons why it was not detected.

The current standard model for GRBs and their afterglows is the fireball shock
model (for a review see M\'esz\'aros 2002). Based on this model, we expect a
prompt multi-wavelength flash, contemporaneous with the gamma-ray emission and
reaching $V$ magnitudes ranging between 9 and 18 for typical GRB distances
(M\'esz\'aros \& Rees 1997; Sari \& Piran 1999a; Sari \& Piran 1999b).  The
prompt optical emission of GRB 990123 is not proportional to the gamma-ray
flux, nor can it be understood as the low-energy extrapolation of the GRB
spectrum (Galama et al. 1999). This indicates that the prompt optical emission
and the GRB originate from different regions in the fireball. It has been
popular practice to ascribe the origin of the GRB to the internal shocks (Rees
\& M\'esz\'aros 1994), and the long-term afterglow to the forward component of
the external shocks. The prompt optical emission is explained as a consequence
of the reverse component of the external shocks (M\'esz\'aros \& Rees 1997;
Sari \& Piran 1999a). The decay rate of the optical flux from the reverse
shocks is much faster than that of the forward shocks, so the emission of the
latter dominates after tens of minutes.  If this interpretation is correct, GRB
990123 and GRB 021211 would be the only two bursts in which all three emitting
regions have been convincingly seen, and demonstrate that afterglow observations may give us
insight into the geometry of the central engine of GRBs, as well as the
structure of the relativistic ejecta.

The interpretation that the prompt and afterglow emissions come from different
physical origins also implies that the two components are largely independent,
so if both are declining as power laws, eq. (1) should be an adequate model to
fit the data. The rather poor fit as discussed in \S 3 is thus surprising.
Besides the possible explanation of a dramatic color change at early times (\S
3), another alternative is that there are small-scale variations in the
reverse-shock emission superimposed on a power-law decline. It is also possible
that the afterglow emission does not begin its power-law decay until after
undergoing a brightening phase to a peak (e.g., Fox et al. 2003), in which case
eq. (1) would not be an accurate model for the data. A full treatment of these
possibilities is beyond the scope of the current {\it Letter}. We also
emphasize that early-time photometry with high temporal resolution, preferably
in multiple passbands, is necessary to determine which explanation is correct.

The small-scale variations in the reverse-shock emission, if real, are
characterized by a timescale of $\Delta t \approx t$. This suggests a process
that involves the whole observable fireball surface.  It will be difficult to
achieve these variations in the reverse shock with external density
inhomogeneities, and a refreshed shock (an extra energy input from the inner
engine) seems to be a more promising explanation (e.g., Panaitescu,
M\'esz\'aros, \& Rees 1998). This would indicate a difference between the
variability in GRB 021211 and GRB 021004, whose variability is believed to be
caused by external density inhomogeneities (e.g., Lazzati et al.  2002).

The OA of GRB 021211 declined 4 mag in half an hour, and was fainter than 20
mag just over 2 hr after the burst. These observations may shed some light on
the mystery of the so-called ``dark bursts," in which no OA is seen (Fynbo et
al. 2001; Berger et al.  2002). Explanations for the non-detections of OAs when
deep, prompt searches exist are that they are (1) obscured by dust in their
host galaxies, (2) at high redshifts, (3) intrinsically faint, and (4) fast
decliners. Out of these, extinction by dust is often favored, since GRBs are
suggested to be associated with massive star formation, and there are GRBs
whose radio and infrared afterglows are detected but not the OA (e.g., GRB
980329; Yost et al. 2002). The relative faintness of the OA of GRB 021211
compared to the OA of GRB 990123, and its rapid optical decay, suggest that
{\it some} dark bursts may be intrinsically dim, and not obscured by dust.  The
faintness of such OAs, coupled with the fast decline typical of OAs, will make
them difficult to observe at high redshifts.  Thus, the diversity in the
underlying properties of OAs must be observationally determined before
substantive inferences can be drawn from the statistics of dark bursts.

The KAIT observations of GRB 021211 demonstrate the importance of rapid
turnaround in localizing GRB positions and intensive followup observations.
Future prospects for optical observations of GRBs are bright: with
continuing contributions from HETE-2, and the planned mission of Swift, fast
and accurate GRB positions will be delivered automatically to the ground for
rapid followup by KAIT and other robotic systems.

\bigskip
\smallskip

We thank S. Barthelmy for his help setting up the KAIT GRB alert system, and
the referee for useful comments. The work of A.V.F.'s group at U. C. Berkeley
is supported by NSF grant AST-9987438.  KAIT was made possible by generous
donations from Sun Microsystems, Inc., the Hewlett-Packard Company, AutoScope
Corporation, Lick Observatory, the NSF, the University of California, and the
Sylvia and Jim Katzman Foundation. S.J. thanks the Miller Institute for Basic
Research in Science at U. C. Berkeley for support through a research
fellowship.

\clearpage
\begin{deluxetable}{cccc|ccc}
%\tablefontsize{\footnotesize}
%\tablewidth{250pt}
\tablecaption{KAIT Photometry of GRB 021211}
\tablehead{
\colhead{$t$ (hr)\tablenotemark{a}} & \colhead{$R$\tablenotemark{b}} &\colhead{$\sigma_R$}&
&\colhead{$t$ (hr)} & \colhead{$R$ mag} &\colhead{$\sigma_R$}
}
\startdata
0.0360 & 14.671\tablenotemark{c} & 0.011& & 0.1533 & 17.066 & 0.047\\
0.0489 & 15.136 & 0.011& & 0.1644 & 17.154 & 0.049\\
0.0614 & 15.540 & 0.011& & 0.1703 & 17.234 & 0.042\\
0.0747 & 15.939\tablenotemark{d} & 0.016& & 0.1758 & 17.248\tablenotemark{d} & 0.038\\
0.0805 & 16.038\tablenotemark{d} & 0.018& & 0.1933 & 17.348\tablenotemark{d} & 0.022\\
0.0861 & 16.117\tablenotemark{d} & 0.019& & 0.2158 & 17.562\tablenotemark{d} & 0.024\\
0.0969 & 16.350\tablenotemark{d} & 0.031& & 0.2542 & 17.760\tablenotemark{d} & 0.021\\
0.1030 & 16.449\tablenotemark{d} & 0.024& & 0.3311 & 18.081\tablenotemark{d} & 0.016\\
0.1086 & 16.555\tablenotemark{d} & 0.050& & 0.4500 & 18.462 & 0.026\\
0.1194 & 16.713\tablenotemark{d} & 0.034& & 0.6017 & 18.828 & 0.178\\
0.1253 & 16.787\tablenotemark{d} & 0.028& & 2.2103 & 20.187 & 0.322\\
0.1311 & 16.925\tablenotemark{d} & 0.043& & 2.3647 & 20.204 & 0.195\\
0.1419 & 16.946\tablenotemark{d} & 0.033& & 2.4100 & 20.087 & 0.246\\
0.1478 & 16.991\tablenotemark{d} & 0.033& & 2.4880 & 20.178 & 0.168\\
\enddata
\tablenotetext{a}{Hours since the burst trigger on 11 Dec. 2002, 11:18:34.03 UT. 
The first four are the flux-weighted mean exposure time using a power-law
decline of index $-$1.6.}
\tablenotetext{b}{Transformed from unfiltered observations; see text.}
\tablenotetext{c}{Superimposed on a longer supernova
followup observation; see text.}
\tablenotetext{d}{Measured on 3 $\times$ 20~s grid images; see text.}
\end{deluxetable}

\clearpage

\begin{figure}
\includegraphics[angle=270,scale=.85]{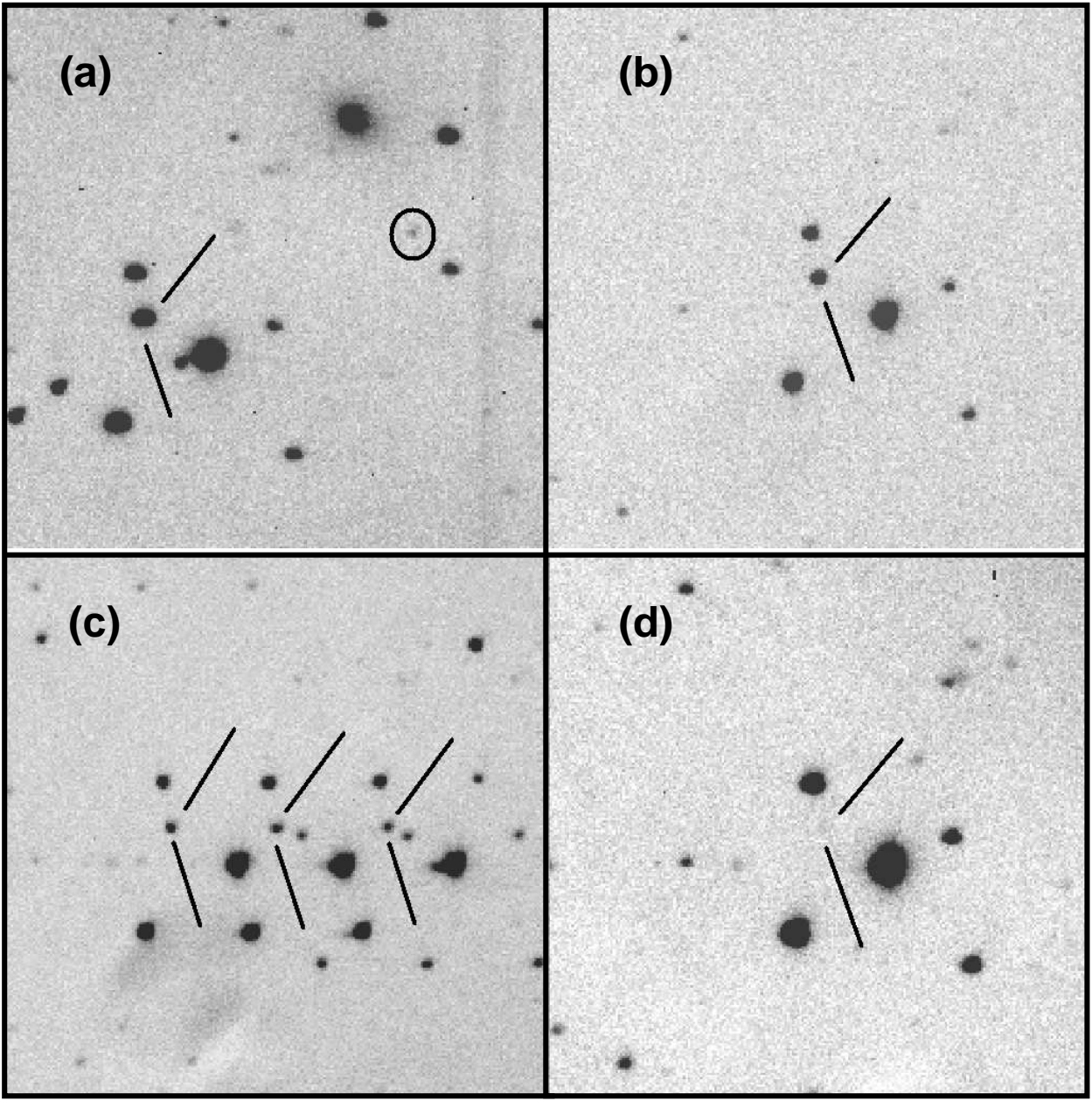}
\caption{KAIT images of GRB 021211. North is up and east to the left. The
field of view in each panel is $3\farcm0\times3\farcm0$. (a) An image 
taken starting at $t = 105$~s; SN 2002he is marked with a circle. (b)
An image taken starting at $t = 171$~s. (c) 
A $3 \times 20$~s grid image taken at $t = 259$~s. (d) 
An image taken at $t = 9600$~s; the position of GRB 021211 is marked with the
crosshairs.}
\end{figure}

\begin{figure}
\includegraphics[angle=270,scale=.75]{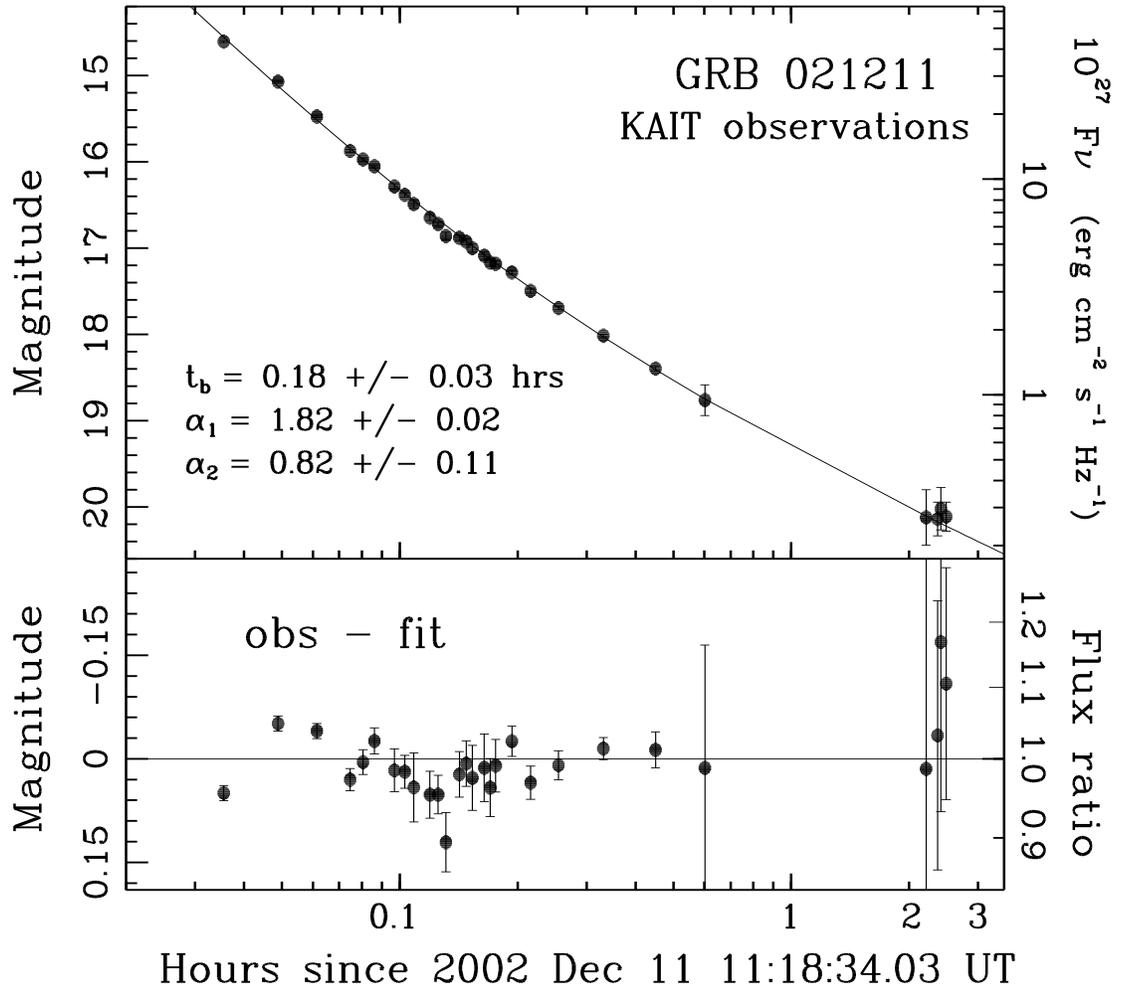}
\caption{Light curve of the OA of GRB 021211 from the KAIT observations.
The upper panel also shows a two-component power-law fit (solid line, see
text for details). The lower panel shows the residuals of the fit.}
\end{figure}

\begin{figure}
\includegraphics[angle=270,scale=.75]{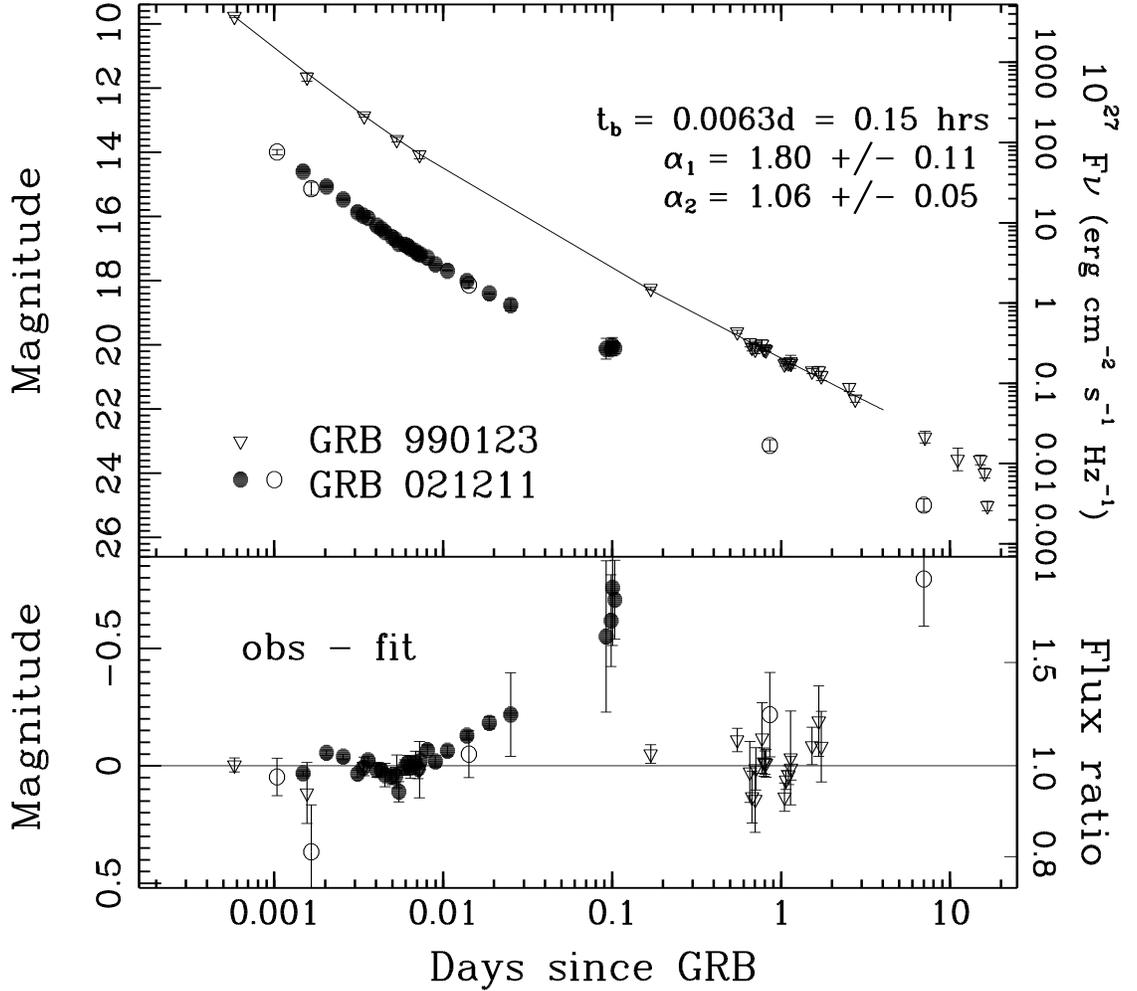}
\caption{Comparison between the $R$-band light curves of 
GRB 021211 and GRB 990123. The solid circles are the KAIT observations
of GRB 021211, while the open circles are those reported in GCN 
Circulars 1757 (Wozniak et al. 2002), 1736 (Park, Williams,
\& Barthelmy 2002), 1750 (McLeod et al. 2002), and 1781 (Fruchter et al. 2002). The solid curve in the upper panel shows the two-component 
power-law fit for the data of GRB 990123 within 2~d after 
the burst. The lower panel shows the residuals of the fit when
the parameters of GRB 990123 are applied to both GRBs. See text for details.}
\end{figure}


\clearpage

\begin{thebibliography}{}
\bibitem[]{392}Akerlof, C. W., et al. 1999, Nature, 398, 400
\bibitem[]{393}Barraud, C., et al. 2003, A\&A, in press (astro-ph/0206380)
\bibitem[]{394}Barthelmy, S. D., \etal~1994, in Proceedings of the 2nd Huntsville 
   Workshop, ed. G. Fishman, J. Brainerd, \& K. Hurley (New York: AIP), 643
\bibitem[]{396}Berger, E., \& Frail, D. A. 2002, GCN Circ. 1745
\bibitem[]{397}Berger, E., \etal~2002, \apj, 581, 981
\bibitem[]{398}Chornock, R., Li, W., Filippenko, A. V., \& Jha, S. 2002, GCN Circ. 1754
\bibitem[]{399}Crew, G., \etal~2002, GCN Circ. 1734
\bibitem[]{400}Filippenko, A. V., Li, W., Treffers, R. R., \& Modjaz, M. 2001, in
Small-Telescope Astronomy on Global Scales, ed. W. P. Chen, \etal~
(San Francisco: ASP), 121
\bibitem[]{403}Fox, D. W. 2002, GCN Circ. 1764
\bibitem[]{404}Fox, D. W., \& Price, P. A. 2002, GCN Circ. 1731
\bibitem[]{405}Fox, D. W., et al. 2003, ApJ, in press (astro-ph/0301377)
\bibitem[]{406}Fruchter, A. S., \etal~1999, \apjl, 519, L13
\bibitem[]{407}Fruchter, A. S., \etal~2002, GCN Circ. 1781
\bibitem[]{408}Fynbo, J. U., \etal~2001, A\&A, 369, 373
\bibitem[]{409}Galama, T., \etal~1999, Nature, 398, 394
\bibitem[]{410}Henden, A. 2002, GCN Circ. 1753
\bibitem[]{411}Hjorth, J., et al. 1999, GCN Circ. 219
\bibitem[]{412}Hoge, J., Willott, C., Grimes, J., Tilanus, R., \& Moriarty-Schieven, G. 2002, GCN Circ. 1742
\bibitem[]{413}Kulkarni, S. R., \etal~1999, \apjl, 522, L97
\bibitem[]{414}Lazzati, D., Rossi, E., Covino, S., Ghisellini, G., \& Malesani, D. 2002, A\&A, 396, L5
\bibitem[]{415}Li, W., \etal~2000, in Cosmic Explosions, ed. S. S. Holt \&
W. W. Zhang (New York: AIP), 103
\bibitem[]{416}Li, W., Filippenko, A. V., Chornock, R., \& Jha, S. 2002, GCN Circ. 1737
\bibitem[]{417}Li, W., Filippenko, A. V., Chornock, R., \& Jha, S. 2003, PASP, submitted
\bibitem[]{418}McEnery, J. 2002, GCN Circ. 1740
\bibitem[]{419}McLeod, B., et al. 2002, GCN Circ. 1750 
\bibitem[]{420}M\'esz\'aros, P. 2002, \araa, 40, 137
\bibitem[]{421}M\'esz\'aros, P., \& Rees, M. J. 1997, \apj, 476, 232
\bibitem[]{422}Panaitescu, A., M\'esz\'aros, P., \& Rees, M. J. 1998, \apj, 503, 314
\bibitem[]{423}Park, H. S., Williams, G., \& Barthelmy, S. 2002, GCN Circ. 1736
\bibitem[]{424}Rees, M. J., \& M\'esz\'aros, P. 1994, \apjl, 430, L93
\bibitem[]{425}Riess, A. G., \etal~1999, \aj, 118, 2675
\bibitem[]{426}Rol, E., \& Strom, R. 2002, GCN Circ. 1777
\bibitem[]{427}Sari, R., \& Piran, P. 1999a, \apjl, 517, L109
\bibitem[]{428}Sari, R., \& Piran, P. 1999b, \apjl, 520, 641
\bibitem[]{429}Schlegel, D. J., Finkbeiner, D. P., \& Davis, M. 1998, \apj, 500, 525
\bibitem[]{430}Vreeswijk, P., Fruchter, A., Hjorth, J., \& Kouveliotou, C. 2003, GCN Circ. 1785
\bibitem[]{431}Wozniak, P., \etal~2002, GCN Circ. 1757
\bibitem[]{432}Yost, S. A., \etal~2002, \apj, 577, 155
\end{thebibliography}
\end{document}